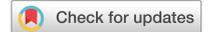

OPEN

# Improved ACOM pattern matching in 4D-STEM through adaptive sub-pixel peak detection and image reconstruction

Nicolas Folastre[1,4], Junhao Cao[1,4], Gozde Oney[1,2,4], Sunkyu Park[1,2], Arash Jamali[1], Christian Masquelier[1,4,5], Laurence Croguennec[2,4,5], Muriel Veron[3], Edgar F. Rauch[3] & Arnaud Demortière[1,4,5] ✉

The technique known as 4D-STEM has recently emerged as a powerful tool for the local characterization of crystalline structures in materials, such as cathode materials for Li-ion batteries or perovskite materials for photovoltaics. However, the use of new detectors optimized for electron diffraction patterns and other advanced techniques requires constant adaptation of methodologies to address the challenges associated with crystalline materials. In this study, we present a novel image-processing method to improve pattern matching in the determination of crystalline orientations and phases. Our approach uses sub-pixel adaptive image processing to register and reconstruct electron diffraction signals in large 4D-STEM datasets. By using adaptive prominence and linear filters, we can improve the quality of the diffraction pattern registration. The resulting data compression rate of $10^3$ is well-suited for the era of big data and provides a significant enhancement in the performance of the entire ACOM data processing method. Our approach is evaluated using dedicated metrics, which demonstrate a high improvement in phase recognition. Several features are extracted from the registered data to map properties such as the spot count, and various virtual dark fields, which are used to enhance the handling of the results maps. Our results demonstrate that this data preparation method not only enhances the quality of the resulting image but also boosts the confidence level in the analysis of the outcomes related to determining crystal orientation and phase. Additionally, it mitigates the impact of user bias that may occur during the application of the method through the manipulation of parameters.

The emergence of new energy materials is related to the development of highly controlled polycrystalline materials exhibiting specific and interesting phase transformation, electronic/ionic conductivity, and optical properties. For instance, knowledge of the spatial distribution of phases, orientations, grain boundaries, and strains is crucial to obtain a complete picture of the phenomena occurring over material operation.

Scanning transmission electron microscopy (STEM) is one of the most developed analytical methods to characterize these polycrystals from the microscopic scale to the atomic scale[1–5]. By coupling the STEM with spectroscopy, hyperspectral data is collected through techniques such as STEM-EDX[6] (energy-dispersive X-ray spectroscopy) for elemental mapping of samples and STEM-EELS[7] (electron energy loss spectroscopy) for chemical environment or oxidation state mappings. In a similar logic, the interest of the 4D-STEM approach is to map structural and chemical information in a 4D image stack (called hyper image) to characterize materials, as the techniques of electron ptychography[8] and real-time integrated center of mass[9]. In the last decade, thanks to new generations of direct electron and hybrid pixel detectors[10], AI computer vision[11], and highly coherent electron beam, a new approach has emerged called 4D-STEM[12–18], in which a large series of diffraction patterns, in near-parallel or convergent beam, are acquired in large stacks of images. Under near-parallel beam conditions (Bragg peaks, spot-like patterns) and using precession, automated crystal orientation mapping (ACOM) turns out to be a new powerful tool to characterize polycrystalline materials at the nanoscale by mapping crystallographic properties[19–21].

[1]Laboratoire de Réactivité et Chimie des Solides (LRCS), CNRS-UPJV UMR 7314, Hub de l'Energie, rue Baudelocque, 80039 Amiens Cedex, France. [2]Institut de Chimie de la Matière Condensée de Bordeaux (ICMCB), Bordeaux, France. [3]Université Grenoble Alpes, CNRS, Grenoble INP, SIMAP, 38000 Grenoble, France. [4]Réseau sur le Stockage Electrochimique de l'Energie (RS2E), CNRS FR 3459, Hub de l'Energie, rue Baudelocque, 80039 Amiens Cedex, France. [5]ALISTORE-European Research Institute, CNRS FR 3104, Hub de l'Energie, rue Baudelocque, 80039 Amiens Cedex, France. ✉email: arnaud.demortiere@cnrs.fr

This 4D-STEM data analysis method based on the ACOM system of NanoMegas (Astar)[22–25] uses pattern matching of a scanning nano-diffraction dataset with libraries of diffraction patterns simulated from known structures extracted from CIF files. This method enables to construct of crystalline phase and orientation maps to determine crystallinity[26,27], microstructures[28], structural deformation[29], and grain boundaries[30] using scanning nano-diffraction with precession mode in a nanometer resolution[31–33].

The recent use of high-speed cameras, pixelated detectors[34] such as CMOS cameras[35], and hybrid-pixel detectors[10] enabled better compromises between signal-over-noise and dwell time of acquisition. However, using such cameras in the column implies strong changes in the acquired images in comparison to the use of a NanoMegas conventional external optical camera, as the quality of the image improves with the increased electron sensitivity and resolution[35]. As the Astar ACOM suite has been optimized for images acquired with the optical camera focused on the phosphorescent screen, the data preparation should be adapted to fit with images acquired using a CMOS camera such as the Oneview Gatan camera. Indeed, as those two camera types present significant differences in dynamic range, resolution and beam sensitivity, the routines should be adapted for two main reasons in this study: (1) to harvest efficiently this increase in signal-over-noise and improve significantly the analysis reliability, (2) to work with lower dose to preserve beam sensitive material and/or achieve faster acquisitions.

The goal of the data preparation methods proposed here is to improve the quality of Astar pattern-matching using a dataset of diffraction patterns acquired with a CMOS Oneview camera. The high sensitivity of the CMOS camera and the data filtering developed here modify the diffraction images leading to a compromise between improving image quality and optimizing template-matching results.

The study utilizes a data reduction technique that employs registration methods to identify electron diffraction spots within patterns. This process enables us to filter and capture the diffraction signal and then, reconstruct the patterns before feeding them into the Astar suite pattern-matching software. We baptized this registration and reconstruction software ePattern. A similar pre-processing workflow is used in the py4DSTEM software package from Savitzky et al.[36] and more recently in the pyXEM[46] library. The essential information of each reflection of a dataset such as intensity, size, and position are recorded in a few minutes with sub-pixel accuracy for the position of the order $10^{-3}$ px, with a data reduction factor of the order $10^2$–$10^3$, meaning the essential information of the diffraction pattern is stored in 100–1000 fewer times space disk. This adaptive method reduces noise and compresses nanodiffraction scanning data for ACOM and strain mapping analysis[37–39], and can also be used on electron diffraction data acquired with other techniques such as 3D electron diffraction (3DED)[40,41].

Firstly, modifications on inside diffraction patterns are estimated through image quality metrics such as peak signal-to-noise (PSNR), structural similarity index measure (SSIM), and root-mean-square error (RMSE). Secondly, the quality of the pattern-matching process on filtered and reconstructed images obtained by the proposed experimental data preparation method is evaluated using index and orientation reliability (OR), as defined in the Astar software. We demonstrate that the experimental data preparation helps to improve the pattern-matching quality result, as it reduces noise overfitting, improves structural similarity index measure, and increases the orientation reliability.

This 4D-STEM study demonstrates the significance of mapping crystal structures and orientations in understanding Na-ion extraction/insertion mechanisms at the individual particles of cathode materials used in Na-ion batteries[42], such as $Na_xMnV(PO_4)_3$ as studied here[27,40,43]. Furthermore, this image processing method, based on a large dataset of electron diffraction, enhances the reliability of phase determinations in complex cathode materials that undergo crystalline transformations and exhibit slight lattice parameter changes.

The beam-sensitive nature of the $Na_xMnV(PO_4)_3$ sample as well as the sensitivity of the camera encourage the use of both lower dose and lower electron flow. The dose used in this study (52 $e^-/Å^2$) gives enough signal to better show and demonstrate the efficiency of the data preparation technique, which can be pushed further. We consider that significantly lower total dose and electron flow can be used with this routine and similar detector sensitivity for more sensitive samples and/or faster acquisitions.

An additional dataset of $PbI_2$ (in liquid) is also used in this study to demonstrate the efficiency of the technique implying different type of noise sources such as the result beam scattering in liquid ($H_2O$). This sample was obtained for a moisture-induced degradation study of CsMAFA to determine the degradation paths of the material along with other techniques.

## 4D-STEM ACOM methods

A 200 kV Tecnai FEI TEM equipped with the Astar system, and a quasi-parallel beam (semi-angle 0.4 mrad) was employed to scan the sample in 10 nm steps using the 4D-STEM technique. To minimize the weight of dynamical effects in the diffraction patterns (DP) and Kikuchi line contrast, precession of the electron beam was utilized, as described in numerous studies[44–46]. Indeed, the dynamic effects induce a change in intensity of the diffraction spot over the entire pattern adding an extra contribution to the diffraction signal, which can be reduced using the beam precession method. As shown in Fig. 1, banks of DPs are simulated and generated for multiple orientations from crystal structure files (CIF) to be compared by cross-correlation to experimental DPs. Optimization is required for various simulation parameters of diffraction patterns, such as the precession angle, spot intensity scale, extension of reciprocal space diffraction figure, and double diffraction conditions, to enable operation.

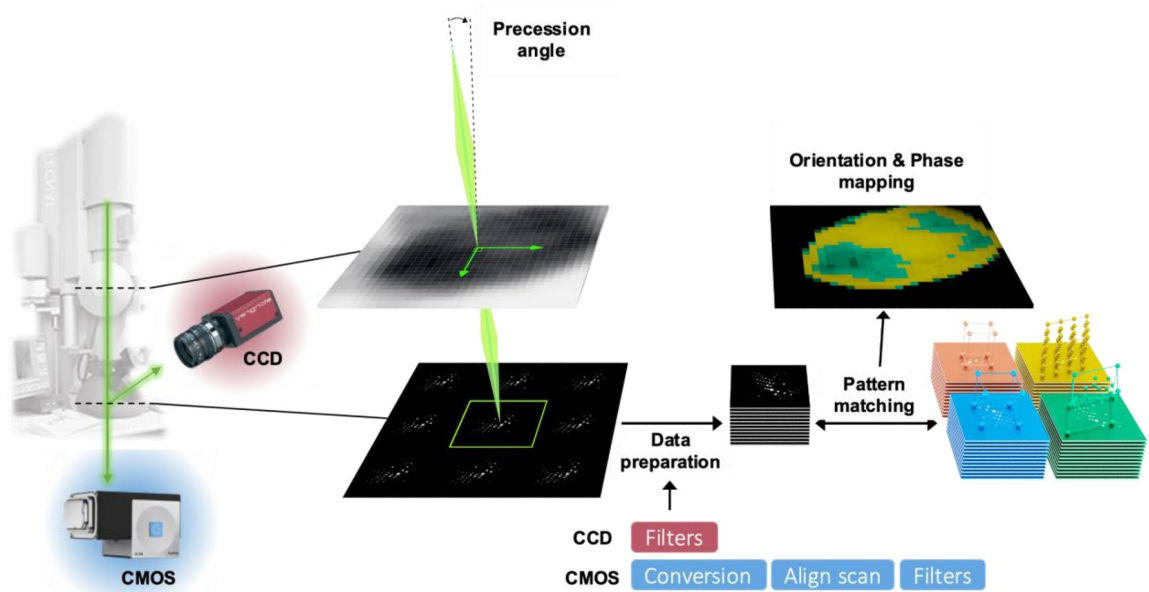

**Figure 1.** 4D-STEM ACOM scheme. Global scheme of 4D-STEM technique. The main steps are (1) the acquisition of diffraction pattern (DP) images with a precessed beam using the Stingray CCD camera or the Oneview CMOS camera in the column, (2) Data preparation, (3) Pattern matching with banks of simulated DPs and, (4) Generation of orientation and phase mapping.

By utilizing a specific range of parameters, the software ensures precise similarity between the experimental and calculated patterns. The user can fine-tune these parameters using the overlay display in the pattern-matching software, selecting the optimal set that corresponds to their data. The calculated points are saved in a bank file that is optimized for performance in pattern-matching calculations. The resulting data is then organized into orientation and crystal phase maps, accompanied by reliability maps that indicate the quality of the pattern-matching results. Overall, this approach assists in determining the orientation and phase of each point on the map among the proposed structures[35]. Given the emergence of new cameras, data preparation methods must be adapted to the use of the Astar suite. Indeed, using a CMOS camera in the column implies strong changes in the acquired images in comparison to the use of a conventional CCD camera. Additionally, users can induce bias in the process by fine-tuning the parameters leading to a dependency on the level of experience in use, which impacts the final results and the related reliability. Minimizing human bias in optimization steps is particularly important for battery materials, in which variations in lithium occupancy can lead to small changes in lattice parameters that are difficult to detect. We have devised a technique for data reduction that addresses these limitations. Our approach involves registration and reconstruction steps, coupled with adaptive prominence, background subtraction, and linear filters, to effectively improve image quality.

### Ethics declarations
We declare that all the methods were performed in accordance with relevant institutional guidelines and regulations.

### ePattern Registration strategy
The detailed steps of our strategy are presented in Fig. 2 and the complete scheme is available in Figure S1. Before registering the diffraction signal, certain preprocessing steps must be undertaken, such as converting data and aligning scans if the camera is unsynchronized. When the scan step is sufficiently small, adjacent images can be summed to increase the signal-to-noise ratio, albeit at the expense of spatial resolution. Astar enables the integration of various dark fields via virtual detectors, such as an annular detector for mapping an amorphous phase. Concurrently, a spot detection technique is utilized to isolate and record the diffraction signal (details in SI). The diffraction signal that is captured can be utilized to mask the filtered input data, or to reconstruct the diffraction patterns in greater detail, devoid of any noise or extraneous components. These patterns play a crucial role in the pattern-matching calculation for orientation and crystalline phase mapping. Additionally, the registered data is employed in creating a spot population map, which can be integrated with various other maps, as illustrated in Fig. 2 and Figure S1. The technique used to separate the registered data belonging to amorphous phases is described in Figure S14.

To prevent overfitting while capturing the position and intensity of reflections, the diffraction pattern needs to be appropriately filtered, accounting for the level of noise present in the image. Initially, the "Rolling ball" method is utilized to eliminate the background from the image, with a diameter larger than the largest spot in the image. This enables the isolation of background information that is sufficiently local, without integrating the diffraction signal peaks. Subsequently, a Gaussian filter is applied to facilitate peak detection. The impact of these filters is displayed in Figs. S5 and S6. The concept of prominence is used to differentiate reflections in the complete signal, as it determines the significance of a peak by taking into account its height compared to the largest neighboring peak.

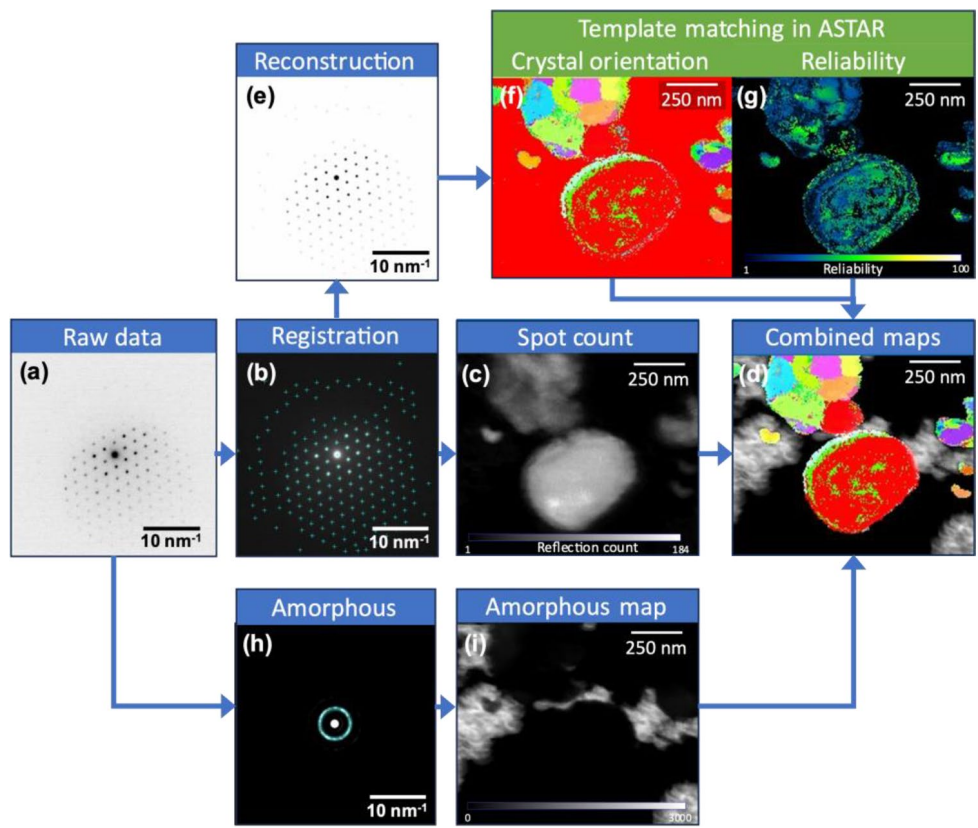

**Figure 2.** ePattern_Registration full block scheme. (**a**) Raw diffraction pattern representing the raw data. (**b**) Registration of spot positions on filtered data. (**c**) Spot count map built from the registration of reflections. (**d**) Combined maps of spot count, crystal orientations, and amorphous phase. (**e**) Reconstruction of diffraction patterns as images using registered data. (**f**) Crystal orientation and (**g**) orientation reliability maps computed in Astar from the reconstructed data. (**h**) Detection of the amorphous phase features and (**i**) corresponding virtual dark field mapping the amorphous phase. The full scheme of ePattern software including data pre-processing and 2D maps combination is presented in Figure S1. The sample is pristine NMVP with black carbon in contact with amorphous carbon.

Once the positions of the peaks are known to the nearest pixel, each reflection position and radius is refined, as shown in Fig. S4. This convergence process starts from the known approximate position and with a radius value smaller than the distance between two spots ideally. The position is refined by calculating the center of mass within its radius, and the radius is simultaneously refined considering that a reflection area with the largest standard deviation contains the whole spot and its immediate surroundings. At the convergence point, the registered radius is 1 pixel smaller than the radius that corresponds to the largest standard deviation. The final position is determined by taking into account the center of mass of several pixel intensities within the final radius. Thus, the position of the reflections of each pattern is determined at a sub-pixel level, whereas the radius is known with one-pixel accuracy. The intensity is computed as the average of the pixels within the spot's radius in the image after removing the background.

Establishing the minimum prominence value required to detect peaks necessitates accounting for the noise level present in the image. Therefore, as illustrated in Fig. 3, the standard deviation is measured at wide angles in the raw image, which directly serves as the minimum prominence value to prevent detecting peaks in the image noise that are not reflections. This adaptive approach to prominence causes a slight loss of information for certain reflections, which becomes more pronounced when detecting peaks at wide angles in the diffraction pattern or when the noise level is high.

### Reconstruction strategy

Our strategy for improving the adaptation of diffraction images for pattern matching involves implementing a data reduction approach that automatically registers essential information in each pattern ($X_{scan}$, $Y_{scan}$), such as the position of reflections ($X_m$, $Y_m$), their radius, and average intensity, as presented in Table 1. The data reduction information (z-latent) obtained from this reflection registration can subsequently be utilized to generate fresh diffraction patterns with a high signal-to-noise ratio. The approach adopts a neural network-like structure consisting of an encoder that extracts the most pertinent features (for registration), a latent space that is the data reduction representation (for parameter table), and a decoder (for reconstruction) to reconstruct the data from the result of latent space but lacks the ability for iterative training.

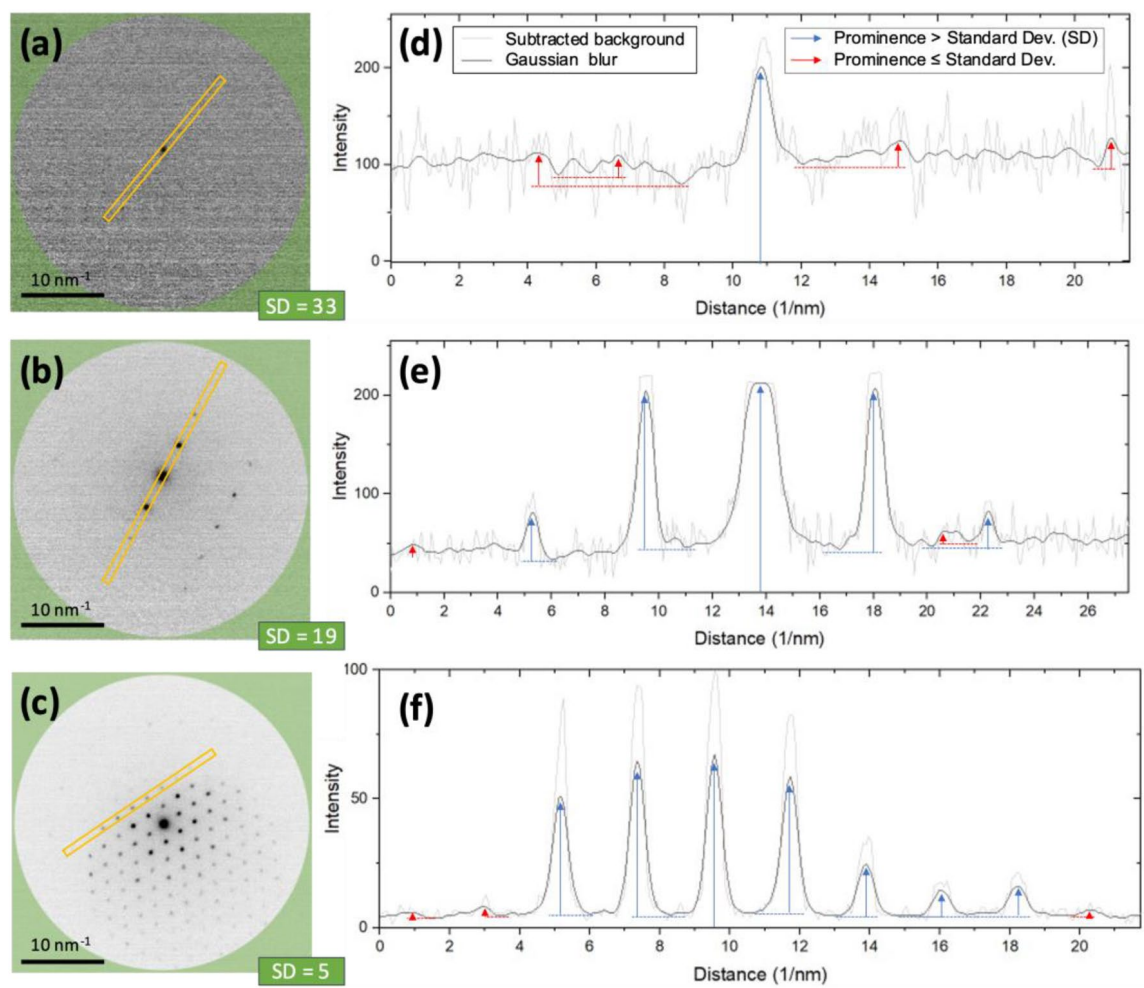

**Figure 3.** Adaptative prominence. Spot detection using adaptative prominence values. (**a**–**c**) Raw diffraction pattern containing different levels of noise. The standard deviation (SD) is measured at high angles in the area represented by the green overlay. (**d**–**f**) Corresponding profiles are integrated into the yellow area. The spots are selected if their prominence is higher than the measured standard deviation.

| Reflection | $X_{scan}$ | $Y_{scan}$ | Xm | Ym | Radius | Mean |
|---|---|---|---|---|---|---|
| 1 | 100 | 100 | 255.992 | 256.105 | 10.000 | 205.750 |
| 2 | 100 | 100 | 203.859 | 259.965 | 9.000 | 96.691 |
| 3 | 100 | 100 | 279.126 | 209.773 | 8.500 | 94.182 |
| 4 | 100 | 100 | 233.237 | 305.041 | 8.500 | 93.116 |
| 5 | 100 | 100 | 282.310 | 267.379 | 5.500 | 120.763 |

**Table 1.** Registration of the first (most prominent) reflections of a diffraction pattern using a subpixel-accurate localization method based on the center of mass (Xm, Ym). The $X_{scan}$ and $Y_{scan}$ denote the position of the pattern in the hyperimage, which is used for the reconstruction, while the $X_m$ and $Y_m$ positions denote the position of the reflection. The radius is determined using an algorithm detecting local standard deviation variation. The mean is the average gray value in this radius. The reflections, in this example, belong to one image whose position is (100,100) in the scan.

The top-right portion of Fig. 4b depicts the usage of a Gaussian filter on the image to enhance the profile of the diffraction peaks and accurately detect their positions. The positions of the peaks, whose prominence exceeds the standard deviation in the raw image at high angles, are recorded to the nearest pixel. In the subsequent step, we refine the precise positions of these peaks by computing their center of mass ($X_m$, $Y_m$), while simultaneously calculating the radius by considering the standard deviation variation around the center of mass.

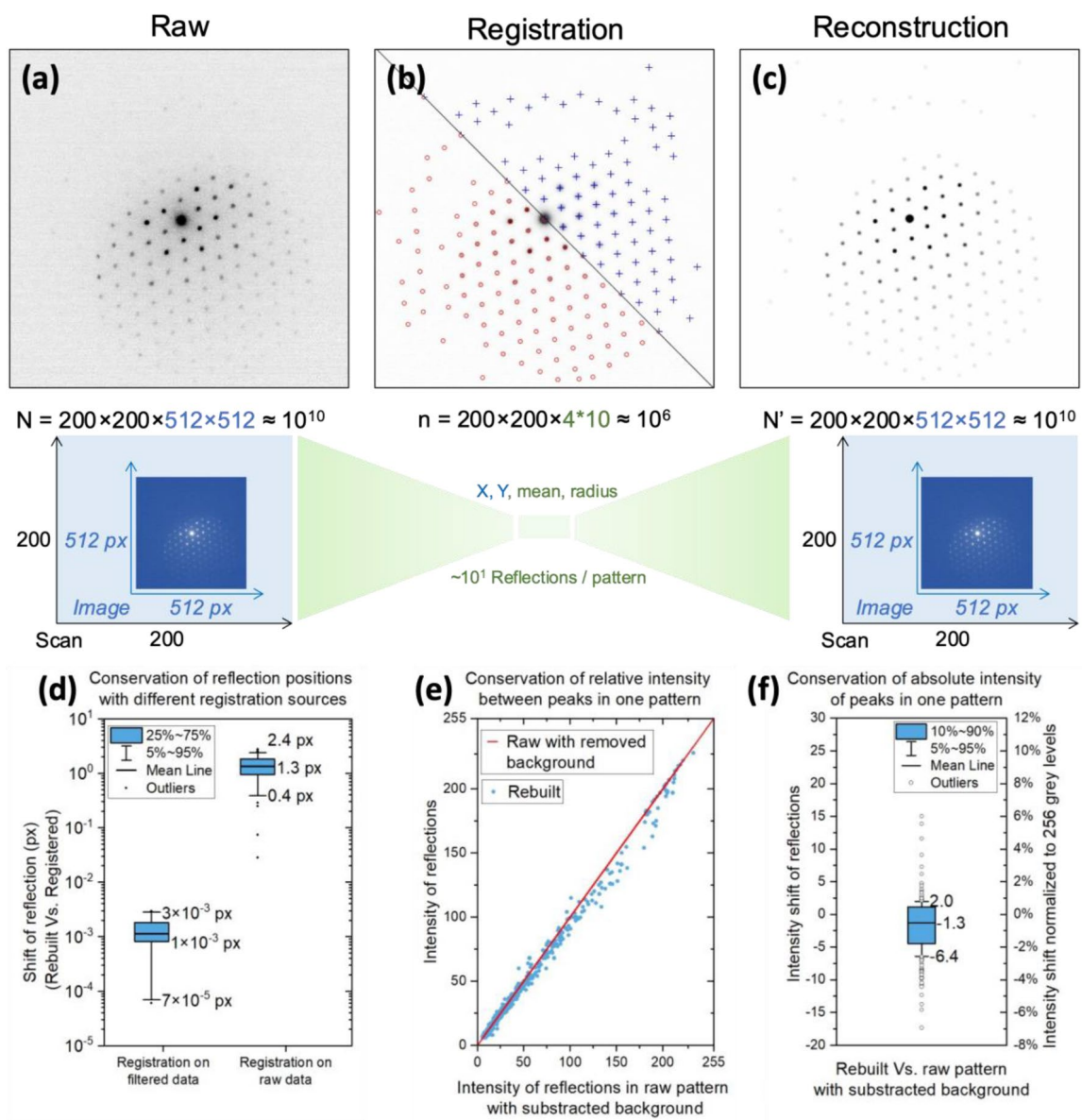

**Figure 4.** Registration and reconstruction. (**a**) the raw diffraction pattern, (**b**) Registration, and (**c**) reconstruction. The top of image (**b**) shows the Gaussian filtering applied to detect the local maxima with pixel accuracy (position spot, blue cross) and the bottom shows the reading of the average intensity of the peaks on the refined positions (intensity spot, red circle). The position, intensity, and size of the reflections of the patterns are recorded in a table which is used to reconstruct the signal in a new so-called reconstructed hyper-image. (**d**) The accuracy of the position of the spots according to the filtering applied beforehand. (**e**) Conservation of relative intensities of peaks in one diffraction pattern between the reconstructed image and the raw image with a subtracted background. (**f**) Difference in intensity between the reflections in the reconstructed image and the raw image without background. The secondary scale shows this difference normalized to 256 levels of gray.

The average gray level within a specified radius is used to measure the intensity of the spot, as depicted in the lower portion of Fig. 4b. The resulting table includes one line for each reflection spot, sorted automatically by significance, as indicated in Table 1. This table displays the most prominent reflections of the same pattern, with the first being the direct beam spot. Figure 4 illustrates the process of registration and reconstruction using this strategy.

The reduction in data size is significant, on the order of $10^3$, as observed in a typical 4D-STEM dataset comprising 40,000 8-bit images of size $512 \times 512$ pixels $(200 \times 200 \times 512 \times 512)$, which are compressed into a table containing 4 parameters for approximately 50 reflections per pattern $(200 \times 200 \times 4 \times 50)$. However, as depicted in Fig. S11, as the number of reflections per pattern increases, the compression rate decreases, whereas fewer reflections per pattern result in a higher compression rate. For the NMVP dataset presented in Figs. 2 and 5, the data reduction factor attained for data written to disk is 614 with an average of 28 reflections per diffraction pattern registered. To go further in detail, there are 10.0 reflections/pattern for the full scan and 28.3 reflection/pattern considering only the patterns with more than 1 reflection (direct beam).

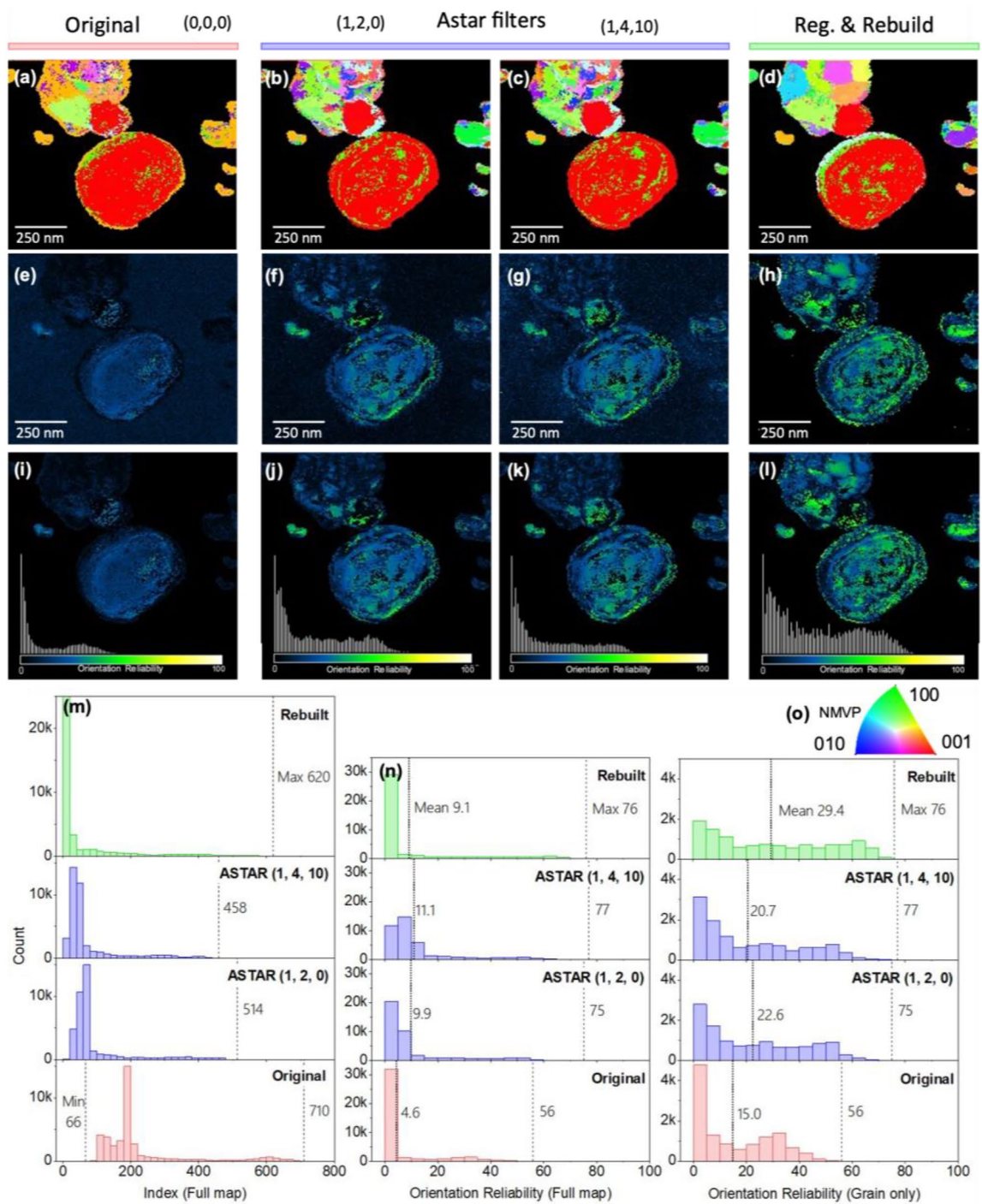

**Figure 5.** Orientation maps of NMVP pristine grains using (**a**) no filter (**b**, **c**) ASTAR filters with 2 different sets of parameters (**d**) Reconstruction using registration of subpixel-accurate positions, radius, and intensity of reflections. The orientation maps are masked to show the grain ROI. Corresponding (**e**–**h**) raw reliability maps (**i**–**l**) masked to show the grain ROI. (**m**) Index of the full maps and (**n**) orientation reliability statistics of the full maps and grains ROI. The grain ROI is defined by a threshold of 5 reflections on the spot count map (Fig. 2). (**o**) Orientation color code for orientation maps.

In the Astar approach, pattern matching involves comparing templates in the form of compressed text files containing points with a stack of diffraction images stored. Therefore, it is necessary to reconstruct the isolated diffraction signal in the form of images from the registered signal. This reconstruction method offers precise and customizable control over parameters such as image scale, reflection radius, and reciprocal radius of the diffraction pattern, depending on the specific study. Figure 4c displays a reconstructed image derived from the registered signal in Fig. 4a, demonstrating a clear background in comparison to the raw image.

The position offset between the original diffraction pattern and the reconstructed one is in the range of $10^{-2}$ to $10^{-3}$ pixels, as illustrated in Fig. 4d. Moreover, registration with the removed background yields a smaller shift in position than with the original image. The center of mass of the spots in the original image depends on the topology of the background, which takes the form of a centered halo, causing the center of mass of the spots to tend towards the center of the image. Thus, to achieve subpixel-level accuracy ($10^{-2}$ to $10^{-3}$ px) in spot position, this registration technique is applied only to images with the background removed.

To achieve subpixel accuracy in controlling the position of an object in an image, an interpolation function is utilized, as shown in Fig. S4. This function adjusts the intensity of pixels in the image and neighboring pixels in two stages (Fig. 4). First, the object is moved to the nearest pixel from its final accurate position, where X and Y are integers. Second, a wider zone is selected around this point, and the interpolated movement occurs by assigning new intensities to the original and neighboring pixels. By redistributing the gray values, this function moves a group of pixels containing and neighboring the object by a subpixel amount along the X and Y directions in the image. It should be noted that this subpixel precision is more useful in motion vector analysis software that requires fine image reconstruction for strain mapping than in pattern-matching software like Astar.

Figure 4e demonstrates that the relative intensity between peaks within the same diffraction pattern is well maintained after reconstruction. The intensity relationship between each peak before and after reconstruction is preserved. However, according to Fig. 4f, some peaks may exhibit absolute intensity variations of up to ±5% of the total scale of gray levels. Moreover, the intensities of the reconstructed image are, on average, lower than those of the original image. This trend might be due to the pixel-level resolution of the spot size, which is determined during the refinement of reflection center positions. If the found radius is not precise enough, the registered averaged intensity might be influenced by darker pixels surrounding the reflections.

The registration and reconstruction process were optimized for speed by opening the data on stacks of images that correspond to a single scan line. The number of iterations required to refine the position of the reflections was reduced, and the reconstruction was done by writing reflections in small stacks in series to limit the amount of RAM required. The masking method mentioned earlier was also implemented to save time. Additional information about the program, functions, and plugins used can be found in the supporting information.

To measure the efficiency of our method, let us consider a stack of 40,000 images with a resolution of 512 px × 512 px. It takes roughly 30 min to register the data with an average of 10 reflections per frame. After registration, the reconstruction from this data takes around 10 min. It is worth noting that we can adjust various parameters such as the size of the final image and the radius of the reflections to enhance the precision of template matching, particularly when differentiating between closely spaced crystalline phases.

The masking method is a time-saving alternative to registration and reconstruction, which sacrifices some precision in spot position (orange block, Fig. S1). Instead of refining the radius or position of the spots, this method quickly identifies their position to mask the rest of the image. This is achieved by modifying pixel values in already-open images before writing them. For a stack of 200 × 200 *images* of 512 × 512 px with about 30 reflections per image, masking can be completed in approximately 20 min.

## Results in ACOM-Astar

The ACOM suite provides orientation and phase maps of crystals, along with associated reliability maps, as a result of calculating cross-correlation. For each point in the scan, the matching index or cross-correlation score is calculated for each orientation of each proposed crystalline phase. This score, denoted by Q, is obtained by summing the products of the points in a pre-calculated diffraction pattern i (known as a template) represented by the function Ti(x, y) and the points in the acquired diffraction pattern represented by the function P(x, y).

$$Q(i) = \frac{\sum_{j=1}^{m} P(x_j, y_j) T_i(x_j, y_j)}{\sqrt{\sum_{j=1}^{m} P^2(x_j, y_j)} \sqrt{\sum_{j=1}^{m} T_i^2(x_j, y_j)}} \tag{1}$$

The highest value of Q provides the orientation solution. The reliability can be calculated by comparing the best index Q1 with the second-best index Q2 using the following equation:

$$R_{orientation} = 100 \cdot \left(1 - \frac{Q_2}{Q_1}\right), Q_1 > Q_2 \tag{2}$$

Greater reliability results in a higher ability to differentiate between two similar orientations on a diffraction pattern, as the ratio between the two highest index solutions is increased.

In Fig. 5, the quality of pattern-matching results is compared for different data preparation strategies: no filters (raw), Astar filters, and the ePattern method presented in this study. The original scan (Fig. 5a,e,i) is indexed without any filter from Astar or external filter. The sample is pristine NMVP in contact with amorphous black carbon.

The Astar filters are:

- Softening loop: the intensity of a given pixel is set equal to the original recorded value minus the average intensity calculated on a circle of radius R centered on that pixel (A 3 by 3 square matrix filter). If negative, the result is set to 0. Increasing the Softening Loops number increases the number of times the filter matrix is applied to the image.
- The "spot enhance": The Spot enhance loops parameter is the number of times the 3 by 3 square matrix filter is applied to the spots of the image. The noise is reduced by the following 3 by 3 square matrix filter applied on every pixel ([1,1,1],[1,8,1],[1,1,1]). Increasing the Spot Enhance Loops number increases the number of times the filter matrix is applied to each of the spots.
- Noise threshold: a threshold of grey values cutting the low intensities. The range is 0–255, and the default value in the software is set to 10. Increasing the Noise threshold will increase the maximum value that sets the intensity values to zero for all pixels whose value is lower than the selected maximum.

The ePattern filters used to extract the features are :

- The Subtract background is a rolling ball strategy that has been chosen for its accuracy rather than the gain in performance compared to the "softening loop" filter. The default range is set to 20 px as the maximum radius of a peak in the dataset.
- Gaussian blur filter of 2 px range by default to (better) detect the positions of the peaks with the 2D prominence algorithm.

Two results using Astar filters with default (Fig. 5b,f,j) and optimized (Fig. 5c,g,k) parameters show the impact of parameter choice on enhancing pattern-matching quality. Finally, the "register and rebuild" result (Fig. 5d,h,l) only utilizes the reconstruction method described in this study, without any Astar filters. The grain ROI is defined by a threshold of 5 reflections on the spot count map, which is intended to isolate the crystalline regions from the rest of the map where we find the amorphous grid membrane. We consider that 3 points out of the central spot is a too low signal to fit a template with our data. The threshold is then set up to $(3 + 1) + 1 = 5$ peaks including the central spot, and this value should be adjusted depending on the proximity of the templates. A comparison of methods to isolate crystalline points of the scan is discussed in Fig. 7.

The orientation maps are shown in Fig. 5a–d to reveal larger and more homogeneous grain domains for the rebuilt signal, in particular on the upper part of the scan. Figure 5e–h demonstrates that our ePattern method outperforms other methods in accurately identifying the structural phase in the crystal.

The index values provide a raw score for pattern-matching and can indicate trends in noise overfitting, especially in regions with no grain, too much diffusion, or poor crystallinity. Lower minimal index values suggest less overfitting. Figure 5m demonstrates that Astar filters and the reconstruction method lead to a significant decrease in noise overfitting, with minimal index values dropping to 0. Furthermore, the distribution of index values shifts towards lower values with Astar filters and even lower with the reconstruction method. This indicates a potential decrease in overfitting, as lower number of reflections are matching with the templates, more likely in the non-crystalline regions. In addition, the comparison between the raw reliability maps (Fig. 5e–h) and those masked by a spot count of 5 (Fig. 5i–l) show non-zero reliability values in non-crystalline zones (amorphous carbon membrane) for raw and filtered data by Astar. In particular, the case of Astar (1, 4, 10) filtering optimized to maximize reliability shows high-reliability values that are close to that crystalline zone of the map. At the same time, applying the mask to the reliability map of the reconstructed dataset has a minimal impact. This point is discussed further in Fig. 7.

The results in Fig. 5n indicate that the mean OR value for the grain ROI is 29.4 using the reconstruction method, which is a 96% improvement compared to no filter (15.0) and a 48% improvement compared to Astar filters (20.7). Furthermore, although the OR values are higher for the reconstruction method than for the Astar filters in the grain ROI, the mean OR value for the entire scan is lower for the reconstruction method (9.1 < 11.1) due to the lower reliability assigned to non-crystalline regions. Assigning a non-zero reliability value to non-crystalline areas would result in an artificially high average OR value for the entire map. It is worth noticing that the optimization of the Astar filter parameters leads to an increase in OR compared to using no filter. However, it is essential to optimize the parameters to effectively reduce overestimated index values caused by noise overfitting and improve OR in the grain region of interest (ROI).

Furthermore, a higher average phase reliability value for the grain ROI indicates greater confidence in the correct matching of crystals with templates. Therefore, a larger difference between the average reliability of the grain ROI and the full map suggests less overestimation of reliability. Figure 5n shows that the ePattern approach significantly reduces overfitting during template matching and that using an unsuitable set of Astar parameters, such as (1,2,0), can cause considerable overfitting in non-crystalline areas compared to using another set of parameters, such as (1,4,10). These observations highlight the effectiveness of various filtering methods in enhancing pattern-matching quality results. Moreover, the ePattern method has the potential to minimize the influence of human bias during the process by adjusting the parameters, which results in independence from the user's level of experience and improves the reliability of the outcomes.

Figure 6 emphasizes the noise-adaptive nature of the registration method. Specifically, sections (g and h) showcase the unprocessed images filtered in Astar and the reconstructed images of various regions of the scan, respectively. These regions feature varying degrees of noise resulting from different phenomena, such as electron beam multi-scattering in the grains (and liquids), which are dependent on their respective thicknesses. Consequently, it is necessary to consider the noise level specific to each diffraction pattern in the scan to adjust the minimum reflection detection threshold, preserve as much diffraction signal as possible from the raw image, and minimize overfitting caused by the noise.

The crystal orientation and reliability maps demonstrate that accounting for noise inhomogeneities in a scan yields higher and more distinct reliability values Fig. 6a–d. This manifests in a sharper distinction between highly crystalline areas and regions with insufficient signal to be included in the calculation. The reliability histograms (Fig. 6e–f) further illustrate a shift towards higher values for the reconstructed diffraction patterns. It is worth noting that this increase is especially beneficial for populations above the reliability threshold of R = 15, as this indicates that the proposed phase and orientation are deemed reliable[20].

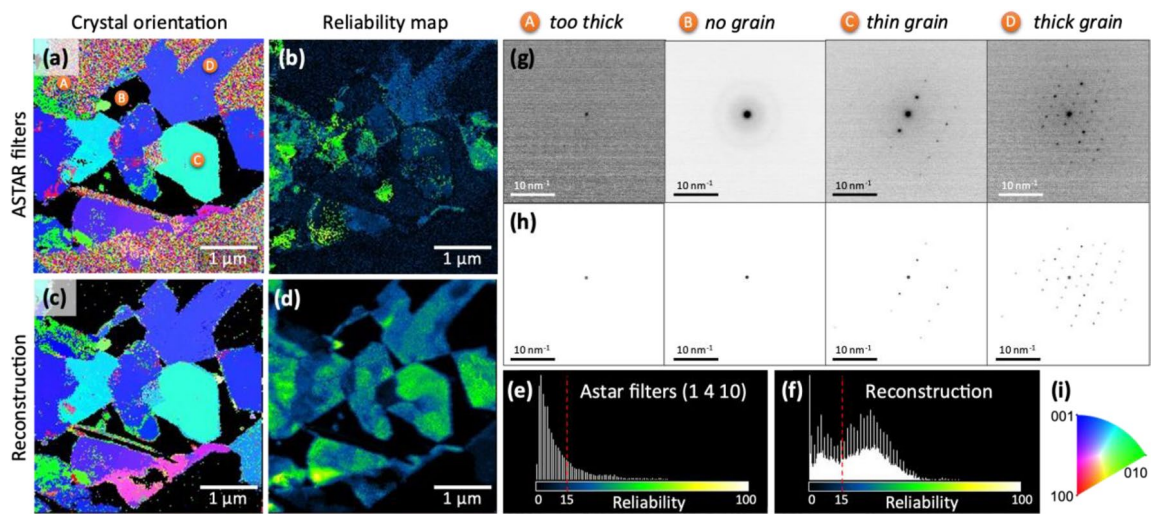

**Figure 6.** Crystal orientation and corresponding reliability maps of $PbI_2$ grains in liquid water (liquid cell TEM holder) using (**a**, **b**) raw images or (**c**, **d**) reconstructed images for pattern matching in Astar. (**e**, **f**) Histogram of corresponding reliability maps. (**g**) Raw diffraction patterns containing different levels of noise and (**h**) corresponding reconstructed images. (**i**) Crystal orientation color code. The sample presented here is $PbI_2$ (in liquid) obtained by water contact induced degradation of CsMAFA.

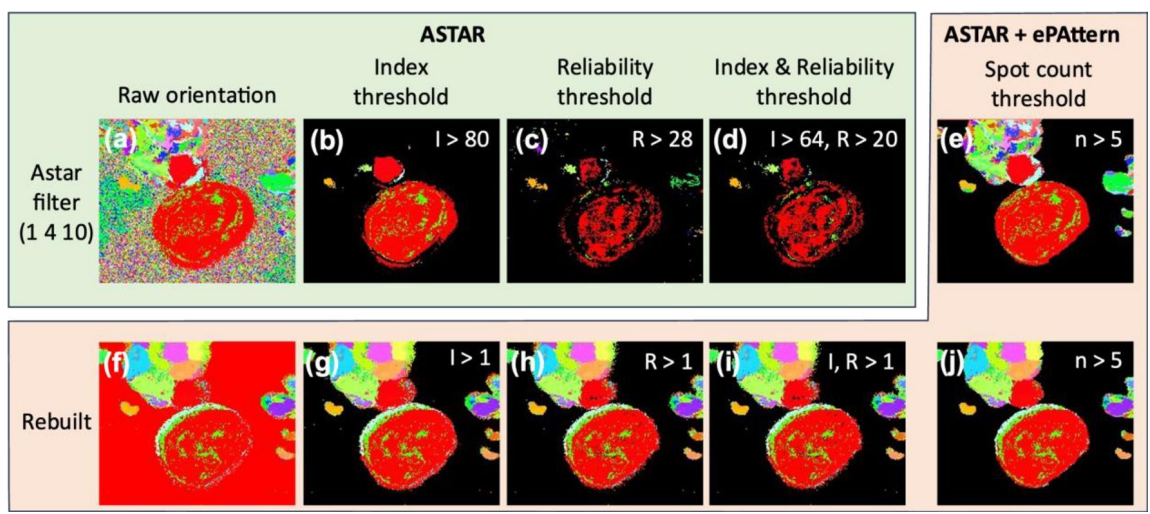

**Figure 7.** Methods to mask the non-crystalline data using Astar and ePattern metrics. (**a**, **f**) Raw orientation maps of two datasets respectively using Astar filters and ePattern reconstruction. (**b**–**d**) Orientation maps of the Astar filtered dataset threshold by the minimum index, reliability, or both values respectively to remove the non-crystalline background. (**e**, **j**) Orientation maps threshold by count of diffraction spots in a DP. (**g**–**i**) Orientation maps of the reconstructed dataset threshold by the minimum index, reliability, or both values respectively to remove the non-crystalline background.

### Handling of result maps

During the analysis of the maps of crystalline orientations and phases, it is advisable to remove the aberrant points not resulting from the diffraction of a crystal. These points are mostly due to template matching of DPs simulated with noise. Indeed, as discussed previously, filtered data may contain artifacts, but the reduction of these artifacts is very efficient by registration and reconstruction of the DPs.

ACOM enables the creation of a virtual darkfield by integrating a portion of the image for each DP in a scan. In this study, this feature was utilized to generate an amorphous phase map, depicted in Fig. 2h,i, and as a means to realign the scan due to the lack of synchronization between the CMOS camera and the beam scan. Simultaneously, the number of spots in an image can be rapidly recorded using a prominence finder algorithm, as shown in Fig. 2b. Consequently, a map is produced by assigning each point a value based on the number of detected spots in each diffraction pattern, as illustrated in Fig. 2c. In the future, this type of map could potentially be utilized to estimate the level of crystallinity or thickness of a sample.

The approach of differentiating between crystalline and amorphous phases provides an opportunity to mask out aberrant points, especially those resulting from artifacts in filtered images. These artifacts are typically situated in areas near the central spot, where the maxima of diffused rings can occasionally register as a small spot.

In the past, a technique employed on unfiltered data involved utilizing an index value threshold to eliminate points where the value is overestimated. However, this threshold method can prove problematic when the noise level is excessively high, and the index value is therefore overestimated. In areas with crystalline components, some points may have a lower index value than others, particularly if the DPs have only a few diffraction spots. Consequently, using this method may lead to the removal of crucial points in an attempt to eliminate outliers. Relying solely on the reliability value to threshold the result is also questionable since an image containing noise and artifacts can yield a high index score for one orientation template and a significantly lower score for others. This scenario results in a strong reliability value at a specific point of the scan which may not be a DP, as it contains no spot.

With the use of the ePattern method, it is now possible to overcome the previously mentioned thresholding issues. Instead of relying on the index value, we can now use the number of registered spots to threshold the resulting map. This approach is feasible since we detect only 1 or 2 "false spots" on the amorphous rings. Thus, it is possible to obtain a map of orientation solutions and crystalline phases by thresholding the points based on the number of recognized reflections in the DP. Figure S12 demonstrates how the orientation reliability is thresholded from the value of 5 which represent the lower limit below which the reliability is low and cannot guarantee the quality of the orientation solution. The representation of the values higher and equal to 15 is set at the same intensity/saturation level as we consider 15 as the upper limit from which the solution is well-assured.

Figure S12 displays the quality of pattern matching after combining the thresholded number of spots map (Fig. 2c) with the index value, which is applicable in the ePattern method. This combination step is crucial to remove aberrant points caused by the previously mentioned artifacts. Generally, using the Astar suite, multiple amorphous phases and crystalline phases can be mapped simultaneously by specifying angle values on the DP. However, if an amorphous phase and a crystalline phase overlap, the cross-correlation result usually displays a higher index for the amorphous phase, requiring careful adjustment of the index coefficient in the software for different phases. In the case of reconstruction, to represent the overlapped crystalline and amorphous phases, it is essential to separate the two results and represent them together using the pattern-matching software on reconstructed images and the virtual dark field on filtered ones, as illustrated in Fig. 2d. The simultaneous representation of amorphous and crystalline phase maps is crucial, for instance, to investigate ion diffusion in grains in carbon-coated active cathode materials, which is particularly important in battery material research.

Figure 7 shows the use of Astar metrics to cut points from the map that do not contain grain. Indeed, as non-crystalline points cannot give relevant results through the pattern-matching calculation, it is necessary to mask them.

In practice, the reliability threshold can generally be set at a recommended level to keep the data reliable, for example at a value between 10 and 20 for the lower threshold and using no upper threshold, and the index threshold is adjusted to cut out the few remaining outliers. However, in the case of this dataset, we notice that the filtering that has been optimized leaves artifacts that lead to the overfitting of non-crystalline points, which results in an increase in the index value at these points. However, as can be seen in Fig. 7b–d, it is necessary to threshold the index value higher than the value of most of the crystalline points, which have a significant number of spots, as shown in the spot count map of Fig. 2. This loss of data thus increases directly with the proportion of artifacts present before pattern matching. In the reconstructed dataset, we notice that the points considered non-crystalline give much lower index and reliability values, as shown in Fig. 7g–i. Thus, we assume that overfitting is much more limited for the reconstructed dataset than for a dataset filtered by Astar by optimizing the filtering parameters. However, from our experience, this hypothesis holds especially when the SNR of the raw pattern is relatively low. In the case of the reconstructed dataset, a threshold by the count of spots makes it possible to refine the mask applied to the pattern-matching results before applying the reliability threshold. However, in the case of a non-reconstructed dataset, the spot count threshold keeps only the grain areas while strongly limiting user interaction with the results as well as the loss of data. Thus, the spot count metric obtained after registration offers a non-crystalline data mask solution whether the dataset is reconstructed or not, as shown in Fig. 7e and j. The use of the spot count followed by the application of a reliability threshold is particularly effective on data with a low SNR and helps to avoid user influence, in particular by using the ePattern reconstruction instead of the Astar filtering. A similar usage of the number of reflections used to separate the crystalline and non-crystalline areas is implemented in the pyXEM library[46,47].

## Conclusions

In this study, a novel sub-pixel adaptive image processing technique, named ePattern, was developed to enhance and optimize pattern-matching for four-dimensional scanning transmission electron microscopy (4D-STEM) data analysis. The ePattern method, adapted to the use of a CMOS camera, successfully isolated the diffraction signal in the image and reduced noise and scattering contributions mainly due to sample thickness, amorphous phases, and lack of crystallinity. Adaptive prominence coupled with background subtraction and linear filters were used to improve image quality. The filtering, registration, and reconstruction techniques used in this study increased the reliability of crystalline orientation and phase maps and, improved the reliability contrast for scans containing regions with different levels of noise. Furthermore, the signal registration compressed the data by a factor of $10^3$, allowing for storage and analysis of very small volumes of data with excellent precision on reflection position parameters.

This study demonstrates that the use of appropriate data preparation techniques can significantly improve the quality of the resulting image and increase confidence in the analysis of outcomes related to determining crystal orientation and phase. Moreover, the conservation of intensity ratios between the diffraction peaks enables dynamic effects in three-dimensional electron diffraction (3DED) reconstruction and automated crystal orientation mapping (ACOM) to be considered. The diffraction data can also be manipulated to quickly map displacements and changes in reflection intensity.

The use of a CMOS camera with an adapted data treatment allowed us to collect and analyze the 4D-STEM data of the two samples using a relatively low dose (52 e-/Å$^2$). Considering the high dynamic range and sensitivity of this type of detector, we consider that similar acquisition with lower dose is achievable without a significant reliability decrease using this type of data preparation.

The ePattern method reduces the impact of human bias during its application through the manipulation of parameters, making it user-independent in terms of experience level and enhancing the dependability of the final results. Future studies may consider an unsupervised strategy (Clustering) upstream of pattern-matching on the Dimensionality reduction using non-negative matrix factorization (NNMF) or Variational AutoEncoder (VAE) to distinguish the relationship or distribution in the extents of different phases and orientations, even if they overlap in the scan. Additionally, further development of the full ACOM method using data reduction and z-latent template-matching may significantly increase cross-correlation performance. Enhancing the reliability and performance of 4D-STEM techniques through these data reduction methods will be beneficial for a wide range of energy material studies, especially those involving closed crystalline structures that are typically associated with high levels of ambiguity.

## Experimental part
### Synthesis

$Na_4MnV(PO_4)_3$ was synthesized using a sol–gel-assisted solid-state reaction method. Firstly, $Na_2CO_3$ (Sigma-Aldrich, 99.5%) and $NH_4H_2PO_4$ (Sigma-Aldrich, 98.5%) were dissolved in deionized water in a molar ratio of 2:3, forming solution A. Meanwhile, $Mn(CH_3COO)_2 \cdot 4H_2O$ (Fluka, 99%), $C_{10}H_{14}O_5V$ (Sigma-Aldrich, 97%), and citric acid (Alfa Aesar, 99%) were dissolved in a mixture of deionized water and ethanol (50/50 by volume) in a molar ratio of 1:1:2, resulting in solution B. Using a magnetic stirrer, solution A was added dropwise into solution B at a constant temperature of 80 °C in an oil bath. The mixture was then dried further in an oven. The resulting powder was heat-treated at 400 °C for 4 h, with a heating rate of 5 °C/min, in Ar atmosphere. Subsequently, the powder was recovered and ground in a mortar, followed by annealing at 800 °C for 10 h, with a heating rate of 2 °C/min, in an Ar atmosphere.

The CsMAFA precursor solution was prepared by dissolving 1.1 mol L$^{-1}$ $PbI_2$ (508 mg), 0.22 mol L$^{-1}$ $PbBr_2$ (80.7 mg), 1 mol L$^{-1}$ FAI (172 mg), 0.20 mol L$^{-1}$ MABr (22.4 mg), and 0.06 mol L$^{-1}$ CsI (17.5 mg) in anhydrous *N*,*N*-dimethylformamide (Sigma-Aldrich) and anhydrous dimethyl sulfoxide (Sigma-Aldrich) with 4:1 ratio (v/v), respectively. The solution was kept at 70 °C for 2 h under stirring and then cooled down to ambient temperature and filtrated through a 0.45-µm syringe filter, yielding a nominal composition of $Cs_{0.05}(FA_{0.83}MA_{0.17})_{0.95}Pb(I_{0.83}Br_{0.17})_3$. The solution was spin-coated according to a two-step procedure inside an Ar-filled glovebox (MBraun Unilab Pro SP), first at 2000 rpm for 10 s and then at 4000 rpm for 20 s. During the second step, 100 µL of chlorobenzene as an anti-solvent was dropped 10 s before the program's end. These films were annealed at 110 °C for 45 min inside the glovebox to crystallize the perovskite structure leading to a mirror-like dark film. This procedure has been optimized to reach the highest power conversion efficiency in full devices, and ca. 20% PCE is obtained typically on average. It leads to films with the same thickness (350 ± 10 nm).

### Sample preparation

The NMVP ($Na_4MnV(PO_4)_3$) sample was prepared on a standard copper grid by dry powder deposition under an Argon atmosphere and mounted on a double-tilt sample holder to be able to orient the grains easily. NMVP belongs to the trigonal r – 3 c space group.

The $PbI_2$ (trigonal P-3m1 space group) sample was obtained in situ by reaction of CsMAFA (perovskite) in water using a Protochip Poseidon sample holder and associated chips to maintain a constant and monitored water flow. The TEM liquid cell is mounted on the TEM sample holder using a binocular and is composed of the chip on which the sample is deposited, a spacer chip, and a gasket. The spacer chip fixes a gap of 1 µm with the sample chip allowing a consequent flow of water to pass, and the gasket between the two chips of the cell ensures the sealing of the sample holder in the TEM column. The lid of the TEM sample holder is screwed on top with a clamping force fixed by a torque screwdriver. The sealing is then tested before the insertion in the TEM with a turbopump. While the sample holder is inserted and under vacuum in the pumping station the pressure must remain invariant, so the plugs of the internal fluidic circuit Inlets and outlets are unscrewed one after the other to check that there is no pressure variation. Then the sample holder is removed and the caps are put back in place. After the sample holder is inserted into the microscope, the plugs of the inlet and outlet are removed. The inlet is connected via a PEEK tubing set to a 5 mL syringe filled with water mounted on a pump, and the outlet is connected to a septum recipe so the whole circuit is sealed.

A flow rate of 5 µL/min. was maintained by the pump (after the characterizations of the pristine material). The arrival of water in the cell was monitored by TEM imaging as it takes several minutes for the water to come to the cell from the syringe through the 50 cm tubing. The thickness of liquid in the cell depends on the spacer that fixes a gap of 1 µm between the chip. However, the difference in pressure between the liquid within the cell (1 atm) and the vacuum in the column of the TEM (10$^{-9}$ Torr) bends the windows outward, increasing the gap between the chips, especially in the middle of the windows.

### Acquisition

FEI Tecnai F20 transmission electron microscope operated at 200 kV in diffraction mode and equipped with a cold FEG and NanoMEGAS Astar system was used for 4D-STEM studies. The 4D-STEM measurements were performed at a camera length set to 490 mm, spot size 5, gun lens 3, and 10 µm condenser aperture (C2). The convergence angle of the electron beam is 0.4 mrad. To reduce the dynamic effects, the electron beam was precessed using the NanoMEGAS DigiSTAR unit (digital precession electron diffraction unit) and TemDPA software with a precession angle of 1.4° and precession frequency of 100 Hz. All scans were performed in a 200 × 200 pixel zone, whereas the width of one pixel ('step width') is 10 nm.

To significantly increase the signal-to-noise ratio in electron diffraction (ED) patterns, the standard AVT Stingray camera supplied with the Astar system was replaced by the Gatan Oneview CMOS camera. The ED patterns were acquired in 512 px with an exposure time of 0.05 s per frame (20 fps). Such a low exposure time was chosen to avoid the sample damage from the electron beam (total scan time did not exceed 33 min) and oversaturation of the camera. Since the Gatan camera was not synchronized with DigiSTAR control through the TemDPA software (NanoMEGAS software package) there was no possibility of ascertaining that the acquisition rate equals the scanning rate, thus, the ED patterns were collected on fly. To overcome this difficulty, the electron beam was blanked at the end of each scan line giving the black end-of-line signatures in the correlation coefficient map. The estimated dose for a single DP is 52 electrons/Å$^2$.

### Image treatment

An overall block scheme is available in Figure S1. Firstly, The diffraction pattern images were converted from 32 bits dm4 to 8 bits bmp with an optimization of the histogram, as the empty levels at the beginning and the end of the histogram are cropped. The individual ED patterns were gathered in the block files using Diffrac2Block software which were further treated in Blockviewer software. The end-of-line signatures appeared as black tortuous continuous lines in a correlation coefficient map. The alignment of the corresponding block files was performed in BlockViewer by use of the local profile of the end-of-line signature and finding of all successive equivalent profiles with the help of a semi-automatic inbuilt procedure.

Then, the images are extracted from the block to be filtered. As any operation combining images of the scan needs both the alignment of the scan and the alignment of diffraction patterns (DPs), the images have been aligned based on the subpixel registration of the central spot position. The reference position is defined as the center of the image (X = Y = 256.00 for 512 px × 512 px image size), and the images are translated using a bilinear interpolation.

### Pattern matching

Simulation of ED template banks, matching of the experimental ED with simulated templates, and generation of the phase maps were performed using DiffGen2, Index2, and MapViewer2 respectively (NanoMEGAS software package). The input parameters specific to pattern matching are the calibration of the image (camera length) in scale and the filtering parameters integrated into the software (value 0 for this study). During the automatic cross-correlation calculation, the point clouds of the templates (orientations) of each bank are superimposed on the experimental images to determine an index score linked to the intensity products between the points and the corresponding aligned pixel.

To rate the quality and reliability of the performed template matching (and, thus, the reliability of the obtained phase maps) two parameters were employed: the cross-correlation index (Qi) and the reliability (R). The first reveals the match between the experimental spot diffraction pattern P(x,y) and the simulated diffraction patterns for all orientations i in a bank of templates Ti(x,y) (Eq. 1).

The highest Qi value is the best match for the given point and is a measure of the agreement between the experimental and simulated pattern. The reliability of that match can be calculated as in Eq. (2) in which $Q_1$ and $Q_2$ are the best match and the second-best match respectively. R values as R < 5 are considered too low, and values as R > 15 as very reliable.

The identification of phases was made by pattern matching using CIF files of the expected pristine $Na_4MnV(PO_4)_3$ (NMVP) and $PbI_2$ structures.

### Computational time

The computational time depends mainly on the CPU cores and graphical processing unit (GPU) configuration and the support where "warm" data is stored (preferably SSD or M2 storage). We have to specify that the method loads small stacks (scan line by line), so the memory should support loading at least a full line of the scan as a stack, and even three lines if the averaging between neighbor DPs is activated. The remaining/elapsed time is displayed by the software and updates periodically. For a stack of 40,000 images with a resolution of 512 px × 512 px, it takes roughly 30 min to register the data with an average of 10 reflections per frame. After registration, the reconstruction from this data takes around 10 min.

### Hardware and software

The Java-FIJI code is run on a PC with Windows OS equipped with Intel Xeon CPU and Quadro P5000 GPUs. The ePattern_Registration developped in this work is based on several FIJI plug-ins and can be downloaded on Git Hub.

### Data availability

The 4DSTEM datasets used and generated in this contribution are available for free download at https://drive.google.com/drive/folders/1_V_V8LjfcLHvkjl8Fy8xWfzVoMl2sHLT.

### Code availability

The ePattern script is available for free download at https://github.com/Nicolas-Folastre/ePattern.

### Acknowledgements

The research presented in this paper has received support from multiple sources. Specifically, funding has been provided by the French Research Agency (ANR) as part of the DestiNa-ion_operando project (ANR-19-CE42-0014). The authors express their appreciation for the proofreading efforts of D. Boursier. Additionally, the UPJV and RS2E electron microscopy platforms were utilized for this research. The authors acknowledge funding from the RS2E network via the STORE-EX Labex Project ANR-10-LABX-76-01. The authors would like to thank Mohammad Ali Akhavan Kazemi and Frédéric Sauvage (LRCS lab) for their assistance in providing photovoltaic perovskite materials and extend their special thanks to the NanoMegas company for their support during the acquisition process. The authors express their gratitude to Fayçal Adrar (LRCS/RS2E) for his invaluable assistance in the testing and comparative analysis of various models in 4D-STEM data processing.

### Author contributions
N.F. and A.D. have conceived the methodology. N.F. made the conceptualization of the code. S.P. has provided battery materials. This work was supervised by A.D., E.F.R. and, L.C. The datasets were acquired and analyzed by N.F., J.C., G.O. A.J and A.D. The ePattern software was coded by N.F. with the validation of E.F.R, M.V., C.M. and N.F. and A.D. wrote the manuscript. All authors participated in the discussion and revision of this paper and finally approved. This work. A.D. was a project administration and funding acquisitor.

### Competing interests
The authors declare no competing interests.

### Additional information
**Supplementary Information** The online version contains supplementary material available at https://doi.org/10.1038/s41598-024-63060-5.

**Correspondence** and requests for materials should be addressed to A.D.

**Reprints and permissions information** is available at www.nature.com/reprints.

**Publisher's note** Springer Nature remains neutral with regard to jurisdictional claims in published maps and institutional affiliations.